\documentclass[12pt,preprint]{aastex}

\usepackage{txfonts}

\shorttitle{Dynamic screening in solar plasma}
\shortauthors{Mao, Mussack, D\"appen}

\begin{document}

\title{Dynamic screening in solar plasma}

\author{Dan Mao}
  \affil{Department of Physics and Astronomy, University of Southern
  California, Los Angeles, California, 90089, USA}

\author{Katie Mussack}
  \affil{Institute of Astronomy, University of Cambridge, Cambridge, CB3 0HA,
  UK} 
  \email{mussack@ast.cam.ac.uk}

\author{Werner D\"appen}
  \affil{Department of Physics and Astronomy, University of Southern
  California, Los Angeles, California, 90089, USA}

\begin{abstract}
In the hot, dense plasma of solar and stellar interiors, Coulomb potentials
are
 screened, resulting in increased nuclear reaction rates.  Although Salpeter's
 approximation for static screening is widely accepted and used in stellar
 modeling, the question of screening in nuclear reactions has been
 revisited.   In particular the issue of dynamic effects has been raised by
 Shaviv and Shaviv who apply the techniques of molecular dynamics to the
 conditions in the Sun's core in order to numerically determine the effect of
 screening.  By directly calculating the motion of ions and electrons due to
 Coulomb interactions, the simulations are used to compute the effect
 of screening  without the mean-field assumption inherent in
 Salpeter's approximation. In  this paper we reproduce their numerical analysis of the screening energy in the plasma of the solar core and conclude that the effects of dynamic
 screening are relevant and should be included when stellar nuclear reaction
 rates are computed.

\end{abstract}


\section{INTRODUCTION}

Under the extreme temperatures and densities of the solar core, the plasma 
is fully ionized. The free electrons and ions interact via the Coulomb 
potential
\begin{equation}
 U(r)=\frac{e^2}{r} \;.
\end{equation}
In this Coulomb system, 
nearby plasma is polarized by each ion. 
When two ions approach with the possibility of engaging in a 
nuclear reaction, each ion is surrounded by a screening cloud. Thus each 
ion is attracted by the electrons and repelled by the protons in its 
partner's cloud. The combined effect 
of the particles in the screening clouds on the potential energy of
the pair of ions is referred to as screening. This screening effect 
reduces the standard Coulomb potential between approaching ions in a 
plasma to a screened potential which includes the contribution to the 
potential from the surrounding plasma. The reduced potential enables the ions
to tunnel through the potential barrier more easily, thus enhancing fusion
rates.  

\citet{Salpeter_1954} derived an expression for the enhancement of 
nuclear reaction rates due to electron screening. By solving the 
Poisson-Boltzmann equation for electrons and ions in a plasma under the 
condition of weak screening ($\phi_{\rm{interaction}}<< k_BT$), Salpeter 
arrived at an expression for the screening energy that is equivalent to 
that of the Debye-H\"uckel theory of dilute solutions of electrolytes 
\citep{Debye_1923},
\begin{equation}
  E_{\rm{screen}} = \frac{e^2}{\lambda_D} 
\end{equation}
where the Debye length, $\lambda_D$, is the characteristic screening length of
a plasma at temperature $T$ with number density $n$, 
\begin{equation}
\lambda_D^2 = \frac{\epsilon_0 k_B T}{ne^2}. 
\end{equation}

Although Salpeter's approximation for screening is widely accepted, 
several papers over the last few decades  \citep[e.g.][]{Shaviv_1996, 
Carraro_1988, Weiss_2001} have questioned either the 
derivation itself or the validity of applying the approximation to hot, 
dense, Coulomb systems like the  plasma of the solar core. Various work 
deriving alternative formulae for \emph{electrostatic} screening
\citep{Carraro_1988, Opher_2000, Shaviv_1996, Savchenko_1999, Lavagno_2000,  
Tsytovich_2000} were refuted in subsequent papers \citep[see][for a summary of
arguments in Salpeter's defense]{Bahcall_2002}. However, the question of
\emph{dynamic} screening remains open. Dynamic screening
arises because the protons in a plasma are much slower than the electrons. They
are therefore not able to rearrange themselves as quickly around individual 
faster moving ions. Since nuclear reactions require energies several 
times the average thermal energy, the ions that are able to engage in 
nuclear reactions in the Sun are such faster moving ions, which
therefore may not be accompanied by their full screening cloud. Salpeter uses
the mean-field approach in which the many-body interactions are reduced to an
average interaction that simplifies calculations. This technique is quite
useful in thermodynamical calculations that rely on the average behavior of
the plasma. However, dynamic effects for the fast-moving, interacting ions lead
to a screened potential that deviates from the average value. The 
nuclear reaction rates therefore differ from those computed with the mean-field
approximation. 

\citet{Shaviv_1996, Shaviv_1997, Shaviv_2000, Shaviv_2001}
used the method  of molecular 
dynamics to model the motion of charges in a plasma under solar 
conditions in order to investigate dynamic screening. The advantage of 
the molecular-dynamics method is that it does not assume a mean field. 
Nor does it assume a long-time average potential for the scattering
of any two charges, which is necessary in the statistical way to solve
Poisson's equation to obtain the mean potential in a plasma. Shaviv and 
Shaviv attribute the differences between their simulations and 
Salpeter's theory to dynamic effects. Since their claims have been met 
with skepticism, we have conducted independent molecular-dynamics 
simulations to confirm the existence of dynamic effects.

\section{NUMERICAL METHODS}

\textbf{For consistency, we developed molecular-dynamics simulations
  using the same basic numerical methods as Shaviv and Shaviv. To
  quell the skeptics, we evaluated the validity of each assumption and
  approximation before implementing it in our own
  simulations. Previous publications \citep{Mao_2004,Mao_thesis,
Mussack_2006, Mussack_2007, Mussack_thesis} describe this process and
show that we did not find any errors in the methods used by Shaviv and
  Shaviv. Once we validated their techniques, we used our
  independently developed simulations to examine dynamic effects in
  screening. 
}

Our numerical simulation consisted of a 3-dimensional box with 1000 
particles (half protons and half electrons) interacting via the Coulomb 
potential. The temperature and density of the solar core ($T=1.6 x 10^7 
\rm{K}$, $\rho = 1.6 x 10^5 \rm{kg/m^3}$) provided the velocity distributions
and inter-particle spacing. We applied periodic boundary conditions and the 
minimum-image convention. In order to deal with the long-range nature 
of the Coulomb potential, we implemented a cut-off radius. With 
this method, particles separated by a greater distance than the cut-off 
radius were not included in the potential sums as explained in section 
\ref{sect:long-range}. Quantum effects were included through the use of 
effective potentials as desribed in section \ref{sect:QM_effects}.  Detailed
analysis of the use of a cutoff radius and effective quantum potential as well
as tests of our simulations can be found elsewhere \citep{Mao_2004,Mao_thesis,
Mussack_2006, Mussack_2007, Mussack_thesis}.

\subsection{Dealing with Long-Range Forces}
\label{sect:long-range}

The long-range nature of the Coulomb potential introduces additional
challenges to a molecular-dynamics simulation. For many of the materials
commonly modeled using molecular dynamics, short-range potentials such as the
Lennard-Jones potential of a simple fluid 
\begin{equation}
 u(r) = 4\epsilon \left[ \left(\frac{\sigma}{r} \right)^{12} - \left(
 \frac{\sigma}{r} \right)^6 \right]
\end{equation}
are appropriate. This potential decreases drastically with the separation of
the two particles. Therefore it can be truncated
quite reasonably at a distance of \(r_{\rm{cut}} = (2.5 - 3) \sigma\). The
Coulomb potential, on the other hand, decays slowly with particle separation. 
Particles at much greater distances must be included in order to accurately
compute the force on each particle. Ideally, the
contribution of each of the infinite image particles should be included in
the potential sums and the potential should never be truncated. Unfortunately
this would take infinite computing time and is therefore
impractical. Fortunately, more efficient techniques have been developed to
compute long-range interactions \citep[see, for example,][]{Frenkel_text,
Allen_text}.  

In their simulations, Shaviv and Shaviv took advantage of the screening of the
plasma to shorten the effective range of the potential. They assume that 
interactions between particles that are farther apart than a few Debye 
lengths will be screened by the surrounding plasma. We have tested the
truncation of the potential and determined that this assumption is appropriate
for our simulation \citep{Mussack_2007, Mussack_thesis}. This enables us to
truncate the interactions beyond a few Debye lengths, reducing the
infinte-range potential to a manageable long-range potential.

\subsection{Quantum Effects}
\label{sect:QM_effects}

When particles in a two-component plasma are closer
 than the thermal deBroglie wavelength
\begin{equation}
 \Lambda_\alpha = \frac{h}{\sqrt{2\pi m_\alpha k_B T}}\:,
\end{equation}
the effects of quantum diffraction and symmetry become significant. The
thermal deBroglie wavelengths for our system are shown in Table
\ref{table:deBroglie_wavelengths}.
\begin{table} 
 \begin{center}
  \begin{tabular}{c c c}
   \hline
     {}  & Mass  & \(\Lambda_{\alpha}\;\) \\
     Particle &  (\(m_e\)) &    \( (a_B)\)\\ \hline \hline
     Proton & 1836 & .008488\\
     Electron & 1 & .3637\\
   \hline  
  \end{tabular}
 \caption{Thermal deBroglie Wavelenghts at $T=1.6 x 10^7 \rm{K}$
   \label{table:deBroglie_wavelengths}   }
 \end{center}
\end{table}

\begin{figure}
\begin{center}
\includegraphics[
     width=150mm,angle=0]{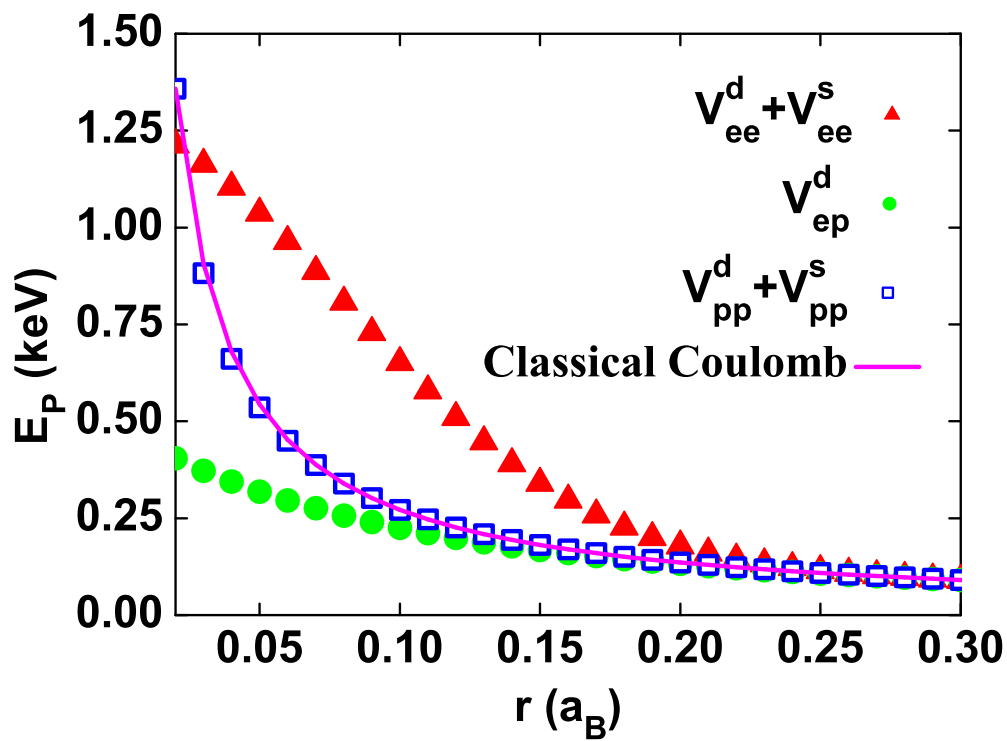}
\end{center}
\caption{Comparison between effective pair potentials including
quantum effects and the classical Coulomb
potential (solid line) in a plasma under the conditions of the solar core.
$V_{ee}$(triangles) and $V_{pp}$(squares) include both diffraction effects
and symmetry effects, while $V_{ep}$(circles) includes only diffraction
effects.} \label{plotone} \vskip 0.5cm
\end{figure}

The average interparticle distance in our model is $<r>= .203 a_B$. Since
 $\Lambda_{p} \ll \; <r>$, a classical treatment is generally sufficient for
 the  protons. However, it will be common to have two electrons that are
 separated 
by a distance less than \(\Lambda_{e}\), requiring quantum mechanical
treatment. We therefore include the quantum approximations used by
\citet{Shaviv_1996}, i.e. the effective pair potentials
derived for a hydrogen plasma by \citet{Barker_1971,Deutsch_1977,Deutsch_1978,
 Deutsch_1979}. Quantum diffraction effects are described by  
\begin{equation}
 v_{\alpha,\beta}^{(d)} =\frac{ e_{\alpha}e_{\beta}}{r}
 \left[1-\exp\left(\frac{-r}{{\lambdabar}_{\alpha,\beta}}\right) \right],
\end{equation}
where 
\begin{equation}
 {\lambdabar}_{\alpha,\beta} = \frac{\hbar}{\sqrt{2\pi\mu_{\alpha,\beta}  
k_B T}}
\end{equation}
 and $\mu_{\alpha,\beta}$ is the reduced mass of the pair.
\begin{equation}
 \mu_{\alpha,\beta} = \frac{1}{\frac{1}{m_\alpha} + \frac{1}{m_\beta}}
\end{equation}
The exclusion principle between electrons is
included by adding 
\begin{equation}
 v_{ee}^{(s)}=k_BT\:\ln 2 \:\exp\left(\frac{-r^2}{\pi{\lambdabar}_{ee}^2\:\ln
 2}\right).   
\end{equation}
 These potentials only differ from the classical Coulomb potential at short
 distances, where quantum effects are relevant. Fig. \ref{plotone} shows
 the contribution to the effective potential from quantum corrections to the
 electrons and protons.  It is clear that quantum
 effects for electrons can not be
 neglected at the high temperature of the solar core. The correction for
 proton-proton pairs is negligible, but electron-proton pairs do experience
 diffraction effects at close distances. Therefore we include quantum
 corrections for both electron-electron and electron-proton interactions.

\subsection{Method}\label{method}

The goal of this work is to compute the energy exchanged between a pair 
of approaching protons and the surrounding plasma in order to evaluate 
the screening energy and the dependence of the screening energy on the 
relative kinetic energy of the pair. In order to accomplish this goal, 
we use a method similar to that of Shaviv and Shaviv. The technique for
tracking approaching proton pairs and the energy they exchange with the plasma
is described below. 

1. For each proton $i$, find all protons approaching it. Choose the 
closest approaching proton and call it the partner of proton $i$.

2. Track the pair as they approach and then move apart. Record their 
distance of closest approach, $r_c$, and the energy of the pair at their 
closest point 
\begin{equation}
 E_{\rm{pair}}^c 
  =E_{\rm{kinetic}}^{c}+\frac{e^2}{r_c}.
\end{equation}

3. Continue tracking the pair until they are separated by a distance 
$R_f$. For our simulations, $R_f=2.0 a_B$ is a sufficient distance to 
represent the separation of the pair. At this point, record the 
far-apart energy of the pair
\begin{equation}
 E_{\rm{pair}}^f
   =E_{\rm{kinetic}}^{f}+\frac{e^2}{R_f}.
\end{equation}
Beyond this point, the pair is no longer tracked.

4. Find a new partner for proton $i$, and repeat the process for $i$ and 
its new partner.

Since a proton pair's departure from their closest point to a distance 
$R_f$ is symmetric to an approach from $R_f$ to $r_c$, we use the 
information obtained from the departure to examine the dynamic 
effect of the plasma on the screening enhancement for interacting proton 
pairs as they approach a separation at which nuclear reactions 
are possible. The dynamic information of each proton and its partner 
\emph{du jour} are recorded at each time step.

\section{RESULTS}

\begin{figure}
\begin{center}
\includegraphics[width=150mm]{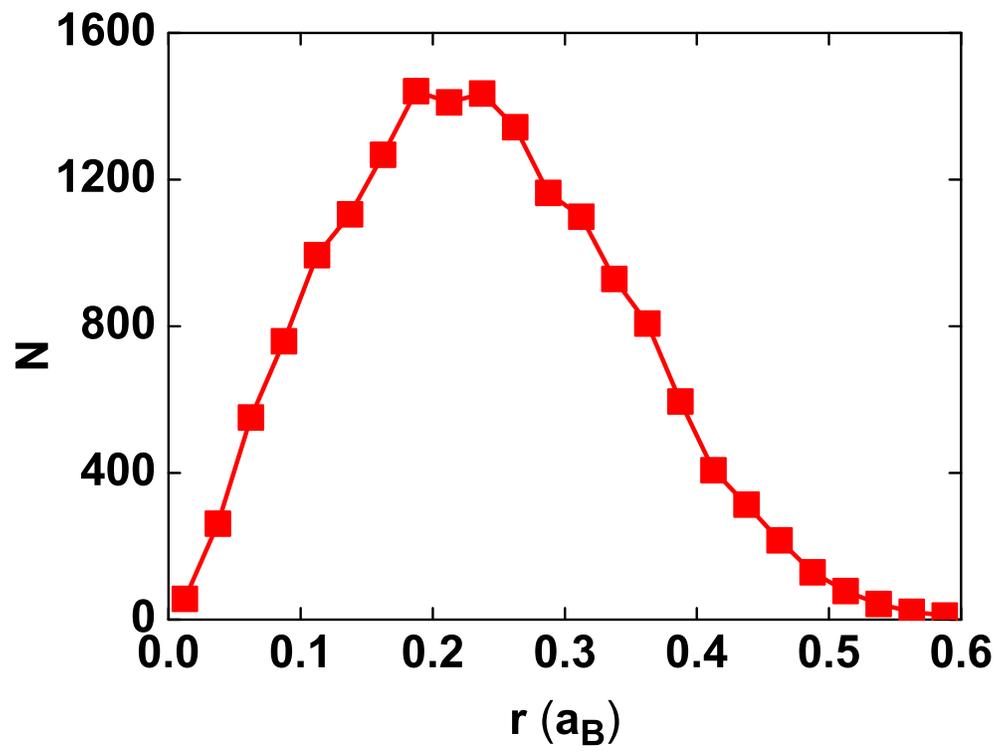}
\end{center}
\vskip -0.2cm \caption{Distribution of the closest distance $r_c$.} 
\label{plottwo} \vskip 0.3cm
\end{figure}

\begin{figure}
\begin{center}
\includegraphics[width=150mm]{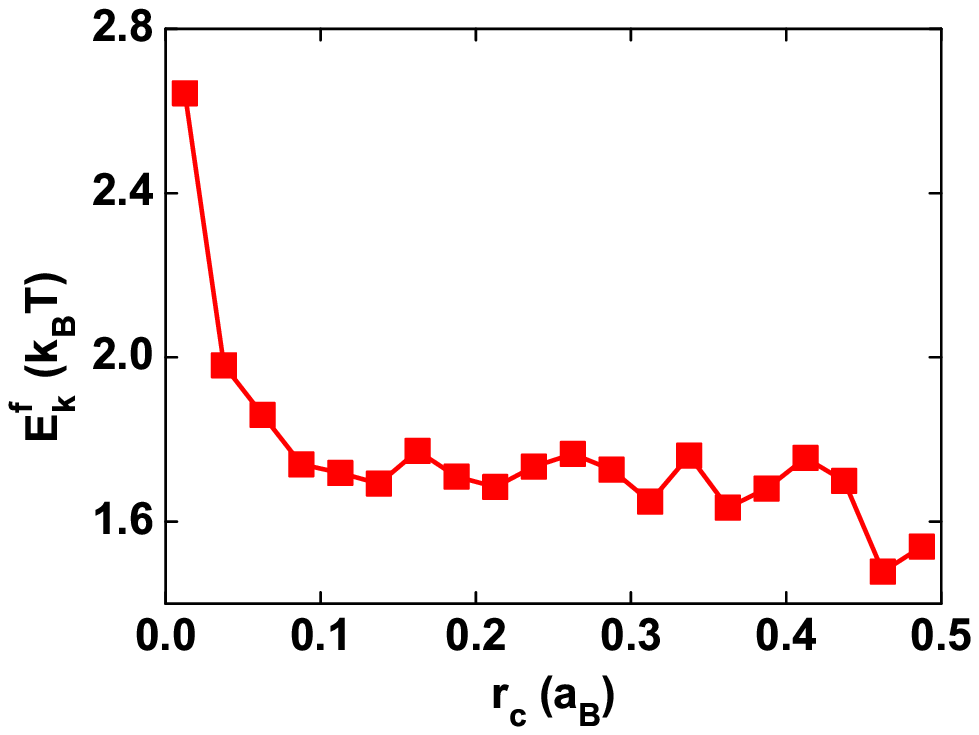}
\end{center}
\caption{The relation between the average kinetic energy of proton
pairs when they are far apart $E_{\rm{kinetic}}^f$ and the closest
separation of the pair $r_c$.}
\label{plotthree}
\end{figure}

First we examine the distribution of the distance of closest approach 
of proton pairs. Fig. \ref{plottwo} shows the number of pairs 
that achieved each closest distance $r_c$. We see that most pairs are 
able to get at least as close as the average distance of neighboring 
protons $<r_p>= 0.256 a_B$.
As seen in Fig. \ref{plotthree}, pairs of protons with higher
far-apart kinetic energy $E^f_{\rm{kinetic}}$ are more likely to get
closer to each other than pairs with lower far-apart kinetic
energy.

The screening energy is the energy exchanged between a pair of protons 
and the surrounding plasma. It represents the combined effect of the 
neighboring protons and electrons on the interacting pair. Salpeter's 
mean-field theory gives an expression for the average screened Coulomb 
potential that includes the energy exchanged between  
interacting protons and the plasma during their approach
\begin{equation}
 \overline{E_{\rm{screen}}}(r)=
  \frac{e^2}{r}\left(1-e^{-r/\lambda_D}\right) \;.
\end{equation}
Salpeter's theory averages the effect of the kinetic energy of each particle
so that the screening energy only depends on distance. However, our
simulations show that the screening energy also depends on the kinetic
energy of the protons. 
Protons with high kinetic energy tend to gain less energy from the plasma
during the interaction, while protons with low kinetic energy gain more
energy from the plasma. \citet{Shaviv_2000, Shaviv_2001} refer to
this as the dynamic effect.

For the purpose of comparison, we define the screening energy in the 
simulations to be the difference in the energy of the pair when they 
are far apart and when they are close together.
\begin{equation}
 E_{\rm{screen}} \equiv E^c_{\rm{pair}} - E^f_{\rm{pair}}
\end{equation}
\begin{equation}
 E^c_{\rm{pair}} = E^c_{\rm{kinetic}} + \frac{e^2}{r_c}
\end{equation}
\begin{equation}
 E^f_{\rm{pair}} = E^f_{\rm{kinetic}} + \frac{e^2}{R_f}.
\end{equation}
Since all pairs have the same far-apart distance $R_f$, 
\begin{equation}
E_{\rm{potential}}^f=\frac{e^2}{R_f}=\mbox{constant}.
\end{equation}
This yields a screening energy of 
\begin{equation}
 E_{\rm{screen}} = E^c_{\rm{kinetic}} - E^f_{\rm{kinetic}} + \frac{e^2}{r_c} -
 E_{\rm{potential}}^f \;. 
\end{equation}

\begin{figure}
\begin{center}
\includegraphics[width=150mm]{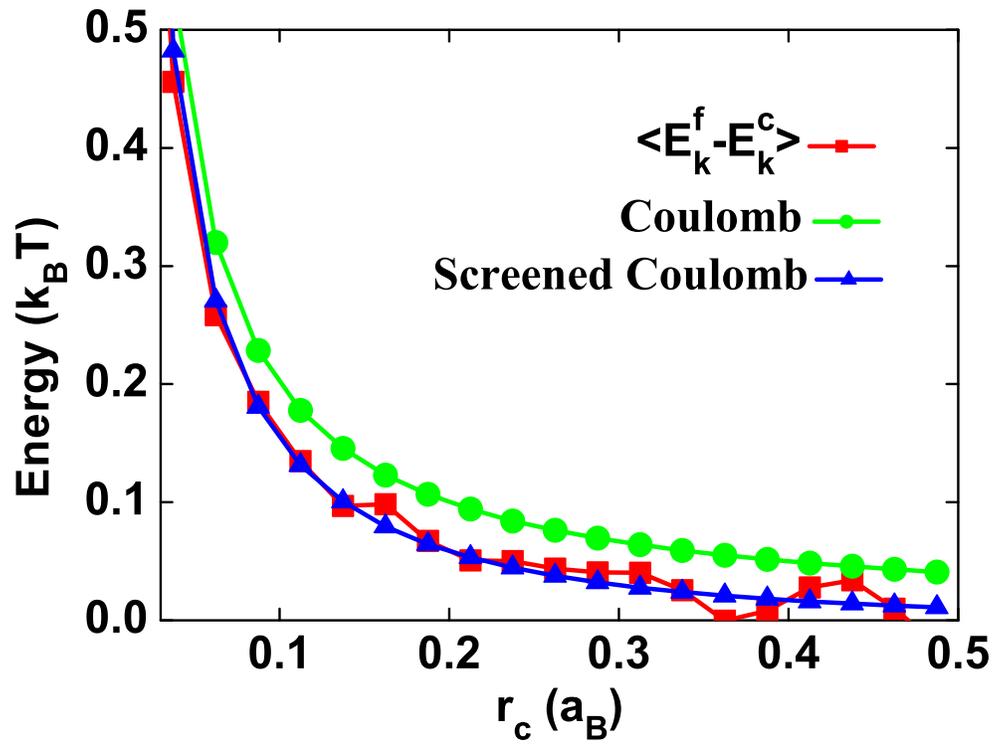}
\end{center}
\vskip -0.4cm \caption{Dependence of the average kinetic energy change
$<\!\!\triangle E_{\rm{kinetic}}\!\!>$ on the closest distance $r_c$, compared
with the Coulomb potential and screened Coulomb potential.}
\label{plotfour} \vskip .3cm
\end{figure}

A key ingredient in this equation for the screening energy is the 
difference in the kinetic energy of a pair when the partner protons are 
far apart and when they are at their closest separation.
\begin{equation}
\!\Delta E_{\rm{kinetic}}\!=(E_{\rm{kinetic}}^f-E_{\rm{kinetic}}^c)
\end{equation}
Here we examine the relationship between the average change in the 
kinetic energy of a pair and their distance of closest approach,  
shown in Fig. \ref{plotfour}. This energy change is compared with 
the Coulomb potential and the screened Coulomb potential from 
Debye-H\"uckel theory. We see that at the distance of closest 
approach, the average kinetic energy exchanged between a pair of 
protons and the surrounding plasma has the form of the screened 
Coulomb potential.
\begin{equation}
 <\!\triangle E_{\rm{kinetic}}\!>\!|_{r_c}\;\sim\,
   \frac{e^2}{r_c}e^{-r_c/\lambda_D}
\end{equation}
On average, the protons interact in a mean field potential of 
\begin{equation}
 V(r)=\frac{e^2}{r}e^{-r/\lambda_D}
\end{equation}
as Salpeter's theory states.

We can rewrite the average screening energy as 
\begin{equation}
 <\!E_{\rm{screen}}\!>\!|_{r_c}\sim
   \frac{e^2}{r_c}\left(1-e^{-r_c/\lambda_D}\right)
   -E_{\rm{potential}}^f.
\end{equation}
As $R_f\rightarrow\infty$, $E_{\rm{potential}}^f\rightarrow 0$, leaving us with 
\begin{equation}
 <\!E_{\rm{screen}}\!>\!|_{r_c}  
  = \overline{E_{\rm{screen}}}(r_c).
\end{equation}
The average screening energy in the simulation is the same as the
average screening energy derived in Salpeter's theory.
This confirms that \emph{on average}, the system can be described by the 
screened Coulomb potential of the Debye-H\"uckel theory as Salpeter 
contends.

\begin{figure}
\begin{center}
\includegraphics[width=150mm]{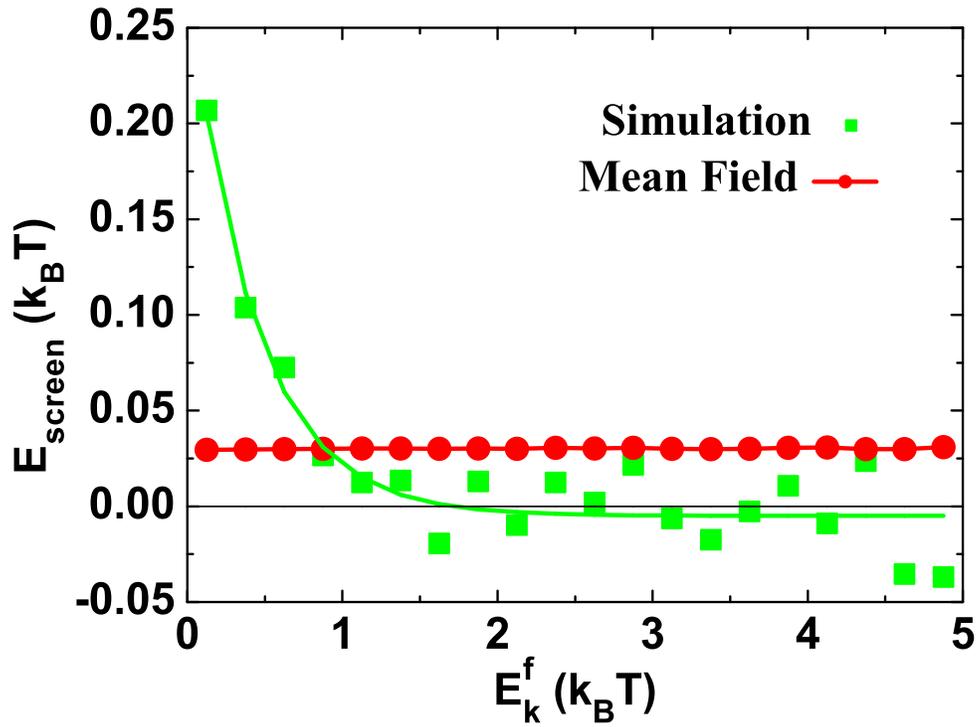}
\end{center}
\vskip -0.4cm \caption{Dependence of the screening energy from the
  simulation 
$E_{\mbox{\scriptsize{screen}}}$ (squares) on the far-apart kinetic energy
$E_{kinetic}^f$. The Debye-H\"uckel screening energy computed at the averege
closest-approach distance of pairs of protons with a given far-apart
kinetic energy is shown (circles) for comparison.}
\label{plotfive} \vskip .3cm
\end{figure}

In order to see the dynamic effect on screening, we must look at the 
relationship between the screening energy and the far-apart kinetic 
energy of proton pairs. Fig. \ref{plotfive} shows the average energy
gained from the plasma by pairs of protons with a given far-apart kinetic energy in
the simulation. For comparison, the Debye-H\"uckel screening energy
computed at a distance equal to the average closest-approach disance
of pairs of protons with each far-apart kinetic energy is also
shown. The screening energy of a pair of interacting protons depends on the relative
kinetic energy of the pair. Pairs with far-apart kinetic energy
greater than the thermal energy tend to gain less energy from the plasma
than the mean-field screening energy, while pairs with lower
far-apart kinetic energy gain energy more energy from the plasma than
the mean-field screening energy.

\begin{equation}
 E_{\rm{screen}} \: > \overline{E_{\rm{screen}}} \:\: for \:\:  E_{\rm{kinetic}}^f < k_BT
\end{equation}
\begin{equation}
 E_{\rm{screen}} \: < \overline{E_{\rm{screen}}} \:\: for  \:\: E_{\rm{kinetic}}^f > k_BT
\end{equation}

When the screening energy from the simulation is averaged over pairs with all
$E^f_{kinetic}$, the result is the same as in Salpeter's theory. This
reaffirms that Salpeter's theory is appropriate as a mean-field
treatment. 

However, the dynamic effect becomes relevant when pairs with high relative kinetic 
energy are singled out, as in solar nuclear reactions. Pairs with
far-apart kinetic energy near the Gamov energy of p+p reactions ($4.8
k_BT$) will, on average, lose energy to the plasma rather than gaining
energy. Because of this dependence of the screening energy on the
relative kinetic energy of a pair of interacting particles, we
conclude that the mean-field approach is not appropriate for
computing nuclear reaction rates.

\section{CONCLUSIONS}

Our simulations confirm that the static Debye screening potential 
describes the average behavior of the plasma as a whole. Therefore it  
remains the appropriate treatment for mean-field properties of the 
system. However, dynamic effects are important when particles are 
singled out based on their velocities. Pairs of particles with greater 
relative velocities experience less screening than pairs of particles 
with lower relative velocities. This effect is relevant in nuclear
reactions in the Sun because they occur at  
energies that are several times greater than the thermal energy. The 
fast protons from the high-end tail of the Maxwell-Boltzmann velocity 
distribution are most likely to take part in these reactions. Thus the 
mean-field approach is not valid in this case. We agree with Shaviv and Shaviv
that the formalism for screening in solar nuclear reaction rates should be
amended to account for this dynamic effect.

\begin{acknowledgements}

We thank Rajiv Kalia, Aiichiro Nakano, and Priya Vashishta for their guidance
on the molecular-dynamics simulations. This work was supported by grant
 AST-0708568 of the National Science Foundation. 

\end{acknowledgements}

\end{document}